\documentclass[10pt]{article}
\usepackage[bottom=22mm,top=19mm,left=20mm, right=20mm]{geometry}

\usepackage{amsmath}
\DeclareMathOperator
\artanh{artanh}

\begin{document}

\title{The Ultimate Limits of the Relativistic Rocket Equation \\
The Planck Photon Rocket}

\author{Espen Gaarder Haug\footnote{e-mail addresses: espenhaug@mac.com.} \\ Norwegian University of Life Sciences, \\ Chr. Magnus Falsens vei 18, 1430 \AA s, Norway}

\date{This is pre-print version 2017 \\  Published in Acta Astronautica, Vol. 136, July 2017, Pages 144-147
 }

\maketitle

\begin{abstract}
In this paper we look at the ultimate limits of a photon propulsion rocket. The maximum velocity for a photon propulsion rocket is just below the speed of light and is a function of the reduced Compton wavelength of the heaviest subatomic particles in the rocket.   We are basically combining the relativistic rocket equation with Haug's new insight on the maximum velocity for anything with rest mass. 

An interesting new finding is that in order to accelerate any subatomic ``fundamental" particle to its maximum velocity, the particle rocket basically needs two Planck masses of initial load. This might sound illogical until one understands that subatomic particles with different masses have different maximum velocities. This can be generalized to large rockets and gives us the maximum theoretical velocity of a fully-efficient and ideal rocket. Further, no additional fuel is needed to accelerate a Planck mass particle to its maximum velocity; this also might sound absurd, but it has a very simple and logical solution that is explained in this paper.


\end{abstract}

\section{Introduction}

Haug \cite{Hau2016n,Hau2016s,Hau2016cc} has recently introduced a new maximum velocity for subatomic particles (anything with rest mass) that is just below the speed of light given by

\begin{equation}
v_{max}=c\sqrt{1-\frac{l_p^2}{\bar{\lambda}^2}}
\end{equation}

where $\bar{\lambda}$ is the reduced Compton wavelength of the particle we are trying to accelerate and $l_p$ is the Planck length, \cite{Pla06}. This maximum velocity puts an upper boundary condition on the kinetic energy, the momentum, and the relativistic mass, as well as on the relativistic Doppler shift in relation to subatomic particles. Basically, no fundamental particle can attain a relativistic mass higher than the Planck mass, and the shortest reduced Compton wavelength we can observe from length contraction is the Planck length. In addition, the maximum frequency is limited to the Planck frequency, the Planck particle mass is invariant, and so is the Planck length (when related to the reduced Compton wavelength).

Here we will combine this equation with the relativistic rocket equation in order to assess how much fuel would be needed to accelerate an ideal particle rocket to its maximum velocity. We will also extend this concept to look at the ultimate velocity limit for a macroscopic rocket traveling under ideal conditions (in a vacuum).

\section{The Limits of the Photon Rocket}

The Ackeret \cite{Ack46} relativistic rocket equation is given by\footnote{See also \cite{Bad53}, \cite{Pom66},  \cite{Vul85}, \cite{Wal06}, \cite{Ant09} and  \cite{Tin07}.} 

\begin{equation}
m_0=m_1\left(\frac{{1+\frac{\Delta v}{c}}}{{1-\frac{\Delta v}{c}}}\right)^{\frac{c}{2I_{sp}}}
\end{equation}

and solved with respect to velocity we have

\begin{eqnarray}
\Delta v &=& c \tanh\left(\frac{I_{sp}}{c}\ln\left(\frac{m_o}{m_1}\right)\right)
\end{eqnarray}

where $I_{SP}$ is the specific impulse, which is a measure of the efficiency of a ``rocket", $m_1$ is the final rest mass of the rocket (payload), and $m_0$ is the initial rest mass of the rocket (payload plus fuel). We will assume that the internal efficiency of the rocket drive is 100 percent, that is $\frac{I_{SP}}{c}=1$. This is basically equivalent to a rocket driven by photon propulsion, or a so-called photon rocket, see \cite{San56,Zelkin60,Sni60,Bur64}. Next we are interested in estimating the amount of fuel needed to accelerate a subatomic particle (using a photon propulsion particle engine) to the Haug maximum velocity, and we get

\begin{eqnarray}
m_0&=&m_1\left(\frac{{1+\frac{\Delta v_{max}}{c}}}{{1-\frac{\Delta v_{max}}{c}}}\right)^{\frac{1}{2}} \nonumber \\
m_0&=&m_1\left(\frac{{1+\frac{c\sqrt{1-\frac{l_p^2}{\bar{\lambda}^2}}}{c}}}{{1-\frac{ c\sqrt{1-\frac{l_p^2}{\bar{\lambda}^2}}}{c}}}\right)^{\frac{1}{2}} \nonumber \\
m_0&=&m_1\left(\frac{{1+\sqrt{1-\frac{l_p^2}{\bar{\lambda}^2}}}}{{1-\sqrt{1-\frac{l_p^2}{\bar{\lambda}^2}}}}\right)^{\frac{1}{2}}  
\end{eqnarray}

when $\bar{\lambda}>>l_p$, as is the case for any observed fundamental particle, we can approximate with a series expansion: $\sqrt{1-\frac{l_p^2}{\bar{\lambda}^2}}\approx 1-\frac{1}{2}\frac{l_p^2}{\bar{\lambda}^2}$, and we get

\begin{eqnarray}
m_0&\approx&m_1\left(\frac{1+1-\frac{1}{2}\frac{l_p^2}{\bar{\lambda}^2}}{{1-1+\frac{1}{2}\frac{l_p^2}{\bar{\lambda}^2}}}\right)^{\frac{1}{2}} \nonumber \\
m_0&\approx&m_1\left(\frac{2-\frac{1}{2}\frac{l_p^2}{\bar{\lambda}^2}}{{\frac{1}{2}\frac{l_p^2}{\bar{\lambda}^2}}}\right)^{\frac{1}{2}} \nonumber \\
m_0&\approx&m_1\left (\frac{4-\frac{l_p^2}{\bar{\lambda}^2} }{\frac{l_p^2}{\bar{\lambda}^2}}\right)^{\frac{1}{2}}  
\end{eqnarray}

Since we assume that $\bar{\lambda}>>l_p$, then this can be further approximated quite well by

\begin{eqnarray}
m_0&\approx&m_1\sqrt{ \frac{4}{\frac{l_p^2}{\bar{\lambda}^2}}} \nonumber \\
m_0 & \approx & m_1 \frac{ 2\bar{ \lambda } }{l_p} =2m_p
\label{rocketmassequation}
\end{eqnarray}

This means that in order to accelerate any particle (an electron, for example) to its maximum velocity we need a particle rocket with two Planck masses of fuel, $2m_p\approx 4.35302\times 10^{-08} \mbox{ kg}$. Appendix A shows a slightly more complex way to derive the same result. The velocity of the electron will then be

\begin{eqnarray}
\Delta v_{max} &=&c \tanh\left(\ln\left(\frac{2m_p}{m_e}\right)\right)=c\frac{\left(\frac{2m_p}{m_e}\right)^2-1}{\left(\frac{2m_p}{m_e}\right)^2+1} \nonumber \\
 &\approx& c\times 0.999999999999999999999999999999999999999999999124  \nonumber \\
\end{eqnarray}

The Einstein relativistic mass\footnote{See \cite{Hau2016n} and  \cite{Ein05}.} of the electron is then equal to the Planck mass.  This is the same maximum velocity as given by \cite{Hau2016n,Hau2016s}. These calculations require very high precision and were calculated in Mathematica.\footnote{We used several different set-ups in Mathematica; here is one of them: \\
\( N[\text{Sqrt}[1-(1616199*10{}^{\wedge}(-41)){}^{\wedge}2\)\(/(3861593*10{}^{\wedge}(-19)){}^{\wedge}2],50]\), where $1616199*10{}^{\wedge}(-41)$ is the Planck length and $3861593*10{}^{\wedge}(-19)$ is the reduced Compton wavelength of the electron. An alternative way to write it is: \\
\({N[\text{Sqrt}[1-(\text{SetPrecision}[1.616199*10{}^{\wedge}(-35)){}^{\wedge}2,50]/(\text{SetPrecision}[3.861593*10{}^{\wedge}(-13)){}^{\wedge}2,50]],50]}\).
}

Remarkably, the concept of two Planck masses being used as fuel to reach the maximum velocity for a subatomic particle holds for any particle. Naturally, this can only work because the maximum velocity of heavier particles is lower than that of lighter particles. According to the concept of the photon rocket, its acceleration is due to the recoil effect in the directed motion of photons. In this case, photons are born as a result of the transformation of particles with nonzero rest mass into electromagnetic radiation.

The equation \ref{rocketmassequation} above is only a good approximation as long as $\bar{\lambda}>>l_p$, which is the case for all observed subatomic particles so far. In the special case, we initially have a payload equal to the Planck mass particle with $m_1=m_p$ we must have  $\bar{\lambda}=l_p$, so we need to use the equation as it was before we applied the series approximation expansion

\begin{eqnarray}
m_0&=&m_1\left(\frac{{1+\sqrt{1-\frac{l_p^2}{\bar{\lambda}^2}}}}{{1-\sqrt{1-\frac{l_p^2}{\bar{\lambda}^2}}}}\right)^{\frac{1}{2}} \nonumber \\
m_0&=&m_p\left(\frac{{1+\sqrt{1-\frac{l_p^2}{l_p^2}}}}{{1-\sqrt{1-\frac{l_p^2}{l_p^2}}}}\right)^{\frac{1}{2}} \nonumber \\
m_0&=&m_p\left(\frac{1}{1}\right)^{\frac{1}{2}} =m_p
\end{eqnarray}

In other words, as we accelerate the Planck mass particle to its maximum velocity we will need no extra mass as fuel. At first this may seem absurd, as we will always need some energy for the acceleration. However, the solution is simple; as \cite{Hau2016n} has shown, the Planck mass particle must always be at rest when observed from any reference frame, the Planck mass particle and the Planck length are remarkably invariant entities. The maximum velocity of a Planck mass particle is 

\begin{equation}
v_{max}=c\sqrt{1-\frac{l_p^2}{\bar{\lambda}^2}}=c\sqrt{1-\frac{l_p^2}{l_p^2}}=0
\end{equation}

The Planck mass particle is the very turning point of light. What is the velocity of light at the precise instant when it changes direction? According to Haug, at this very instant it will be at rest. In the very next instant, the Planck particle will likely be dissolved into energy.

\section{The Maximum Velocity of a Rocket Ship}

The maximum velocity of any composite object (even a nucleus) is likely to be limited by the fundamental particle with the shortest reduced Compton wavelength from which it is constructed. In other words, the speed limit of a rocket is limited by the heaviest subatomic ``fundamental" particle it is built from. When this particle reaches its maximum velocity, that is given by 

\begin{equation}
v_{max}=c\sqrt{1-\frac{l_p^2}{\bar{\lambda}^2}}
\end{equation}

it will first turn into a Planck mass particle and will then likely burst into energy. If this type of fundamental particle is a significant part of the macroscopic object (spaceship) we are traveling in, then the whole ship is likely to be destroyed at the moment we reach this velocity. If a proton was a fundamental particle, then the maximum velocity of a rocket traveling under ideal conditions (in a vacuum) would be\footnote{Here assuming $l_p=1.616199 \times 10^{-35}$ and $\bar{\lambda}_P=2.10309 \times 10^{-16}$.}  

\begin{eqnarray}
v&=&c\sqrt{1-\frac{l_p^2}{\bar{\lambda}_P^2}} = c\times 0.99999999999999999999999999999999999999705  \nonumber \\
\end{eqnarray}

For comparison, at the Large Hadron Collider in 2008, the team talked about the possibility of accelerating protons to the speed of 99.9999991\% of the speed of light \cite{Bru08}. When the Large Hadron Collider went full force in 2015, they increased the maximum speed slightly above this (likely to around 99.99999974\% of the speed of light).  In reality, if a proton consists of a series of other subatomic particles, then the speed limit given above for a proton would not be very accurate. Alternatively, we could have looked at the reduced Compton wavelength of the quarks that the standard model claims make up the proton.

\section{Summary and Conclusion}

The maximum amount of fuel needed for any fully-efficient particle rocket is equal to two Planck masses. This amount of fuel will bring any subatomic particle up to its maximum velocity. At this maximum velocity the subatomic particle will itself turn into a Planck mass particle and likely will explode into energy. Interestingly, we need no fuel to accelerate a fundamental particle that has a rest-mass equal to Planck mass up to its maximum velocity. This is because the maximum velocity of a Planck mass particle is zero as observed from any reference frame. However, the Planck mass particle can only be at rest for an instant. The Planck mass particle can be seen as the very turning point of two light particles; it exists when two light particles collide.\footnote{See \cite{Hau2016n} for a detailed discussion and presentation of an atomism particle model.} Haug's newly-introduced maximum mass velocity equation seems to be fully consistent with application to the relativistic rocket equation and it gives an important new insight into the ultimate limit of fully-efficient particle rockets.

\section*{Appendix A}

This shows a slightly different and slightly more complex way to derive the same result. In the case of a photon rocket, when combined with Haug's maximum velocity for subatomic particles, we have

\begin{eqnarray}
\Delta v_{max} &=& c \tanh\left(\ln\left(\frac{m_o}{m_1}\right)\right) \nonumber \\
\frac{\Delta v_{max}}{c} &=& \tanh\left(\ln\left(\frac{m_o}{m_1}\right)\right) \nonumber \\
\artanh\left(\frac{\Delta v_{max}}{c}\right) &=& \ln\left(\frac{m_o}{m_1}\right) \nonumber \\
e^{\artanh\left(\frac{\Delta v_{max}}{c}\right)} &=& \frac{m_o}{m_1} \nonumber \\
m_0&=&m_1e^{\artanh\left(\frac{\Delta v_{max}}{c}\right)} \nonumber \\
m_0&=&m_1e^{\artanh\left(\frac{c\sqrt{1-\frac{l_p^2}{\bar{\lambda}^2}}}{c}\right)} \nonumber \\
m_0&=&m_1e^{\artanh\left(\sqrt{1-\frac{l_p^2}{\bar{\lambda}^2}}\right)}  \nonumber \\
m_0&=&m_1e^{\frac{1}{2}\ln\left(\frac{1+\sqrt{1-\frac{l_p^2}{\bar{\lambda}^2}}}{1-\sqrt{1-\frac{l_p^2}{\bar{\lambda}^2}}}\right)}  
\end{eqnarray}

Further, when $\bar{\lambda}>>l_p$ we can use a series approximation, $\sqrt{1-\frac{l_p^2}{\bar{\lambda}^2}}\approx  1-\frac{1}{2}\frac{l_p^2}{\bar{\lambda}^2}$, this gives

\begin{eqnarray}
m_0&\approx &m_1e^{ \frac{1}{2}\ln\left(\frac{1+1-\frac{1}{2}\frac{l_p^2}{\bar{\lambda}^2} }{1-1+\frac{1}{2}\frac{l_p^2}{\bar{\lambda}^2}}\right)} \nonumber \\
m_0&\approx&m_1e^{\frac{1}{2}\ln\left(\frac{2-\frac{1}{2}\frac{l_p^2}{\bar{\lambda}^2}}{\frac{1}{2}\frac{l_p^2}{\bar{\lambda}^2}}\right)} \nonumber \\
m_0&\approx&m_1\left(\frac{2-\frac{1}{2}\frac{l_p^2}{\bar{\lambda}^2}}{\frac{1}{2}\frac{l_p^2}{\bar{\lambda}^2}}\right)^{\frac{1}{2}} \nonumber \\
m_0&\approx&m_1\sqrt{\frac{4-\frac{l_p^2}{\bar{\lambda}^2}}{\frac{l_p^2}{\bar{\lambda}^2}}}  
\end{eqnarray}

Further, when $\bar{\lambda}>>l_p$, then then this can be very well-approximated by

\begin{eqnarray}
m_0&\approx& m_1\sqrt{\frac{4}{\frac{l_p^2}{\bar{\lambda}^2}}} \nonumber \\
m_0&\approx& m_1\frac{2\bar{\lambda}}{l_p}=2m_p
\end{eqnarray}

This is the same result as we obtained in the main part of the paper using a slightly easier derivation.

 \section*{Acknowledgments}

 Thanks to Victoria Terces for helping me edit this manuscript. Also thanks to Alan Lewis, Daniel Duffy, ppauper, and AvT for useful tips on how to do high precision calculations. Also thanks to an anonymous referee for useful comments.

\bibliographystyle{unsrt} 

\end{document}